\documentclass[aip,apl,twocolumn,floatfix,showpacs,showkeys]{revtex4}
\usepackage[pdftex]{graphicx}
\usepackage{bm}
\usepackage{xcolor}
\usepackage{wasysym}
\usepackage{mathrsfs}
\usepackage{fancyhdr}

\renewcommand{\deg}{$^{\circ}$\hspace{1mm}}
\newcommand{\etal}{{\it et al.}}
\newcommand{\newc}{\newcommand}
 
\newc{\be}{\begin{equation}}
\newc{\ee}{\end{equation}}
\newc{\bfe}{\begin{floatequation}}
\newc{\efe}{\end{floatequation}}
\newc{\bea}{\begin{eqnarray}}
\newc{\eea}{\end{eqnarray}}
\newc{\ie}{{\it i.e.}}
\newc{\eg}{{\it e.g.}}
\newc{\etc}{{\it etc.}}
\newc{\ra}{\rightarrow}
\newc{\lra}{\leftrightarrow}
\newc{\lsim}{\buildrel{<}\over{\sim}}
\newc{\gsim}{\buildrel{>}\over{\sim}}

\begin{document}

\title{Electronic control of magnonic and spintronic devices} 

\author{C. Tannous}
\affiliation{Laboratoire de Magn\'etisme de Bretagne, UBO CNRS-FRE 3117, 6 Avenue le Gorgeu C.S.93837,
29238 Brest Cedex 3, FRANCE.}

\author{J. Gieraltowski}
\affiliation{Laboratoire des Domaines Oc\'eaniques, IUEM CNRS-UMR 6538, 
Technop\^ole Brest Iroise, 29280 Plouzan\'e, FRANCE.}

\begin{abstract}
Nanometric magnonic and spintronic devices need magnetic field control in addition to
conventional electronic control. In this work we review ways to replace
magnetic field control by an electronic one in order to circumvent
appearance of stray magnetic fields or the difficulty of creating 
large magnetic fields over nanometric distances. Voltage control 
is compared to current control and corresponding devices are compared from their 
energetic efficiency point of view.
\end{abstract}



\maketitle

\textcolor{blue}{Version: \today}

\section{Introduction}
Minimal feature in microelectronics CMOS planar technology progressed from 22 nm in 2012
to its present value of 14 nm  and is slated to reach 10 nm in 2016. 

Progressing toward nanoelectronics with a minimal feature approaching
steadily the nanometer limit leads to at least five major consequences:

\begin{enumerate}
\item Joule effect increase.
\item Interconnection delay increase. 
\item Circuit size decrease with respect to some reference electromagnetic (EM) wavelength. 
\item Enhancement of quantum effects.
\item Emergence of spin degrees of freedom.
\end{enumerate}

Joule effect enhancement is due to resistor $R$  scaling by a factor $s > 1$ increasing the 
resistance to $sR$ whereas wire (interconnection) delay 
increase is due to rise of communicating wire length with number of components (see Table~\ref{tab1}).

\begin{widetext}
\begin{center}
\begin{table}[htbp]
\begin{tabular}{|l|c|c|c|c|c|c|}
\hline
Microprocessor (year) & Feature Size  & Transistors & Frequency & Pins  & Power consumption & Wire Delay  \\
 & (nm) & & (MHz) &  & (Watts) &  (clock cycle/cm) \\
\hline
Intel 4004 (1971) &  10,000 & 2,300 & 0.750 & 16 & 1 & 1/1,000 \\
\hline 
Pentium IV-670 (2005) & 90 & 169 million & 3,800 & 775 & 115 &  4 \\
\hline
(2005)/(1971) Ratio & (1/111) & 73,478 & 5,066 & 48 & 115 & 4,000 \\
\hline
\end{tabular}
\caption{Comparison of two CPU characteristics dating from 1971 and 2005 and the evolution 
of their intrinsic properties with feature size decrease. The 4004 is the first 
commercial 4-bit CPU whereas the Pentium IV-670 is a single-core 64-bit CPU. 
As number of transistors, frequency, number of pins all increase as expected, 
wire delay and power consumption increase against all expectations.  
Wire delay is extracted from the speed of an electronic signal which is about 20 cm per 
nanosecond (ns) within the wire. Hence one gets 0.5 ns for a 10 cm propagation length and in the case of
a 1 GHz CPU, the delay represents half-a-clock cycle.}
\label{tab1}
\end{table}
\end{center}
\end{widetext}

While Joule effect and interconnection delay have been somehow 
virtually solved in microprocessors by introducing 
multi-core architecture (see Table~\ref{tab1}), the latter concept appears
as an intermediate solution awaiting a better technology to address, 
in a deeper way, power consumption and wire delays in devices. 

Power consumption issue is tackled in magnonic and spintronic devices 
that are based respectively on spin-wave 
and spin current manipulation.  Extremely low power
dissipation is reached in both cases (except with spin-transfer torque 
(STT) currents where critical density required for switching is very 
large as described in next section) since no Joule effect is expected 
when no charges are moving in order to carry information. 
Undulation of spin moments is the carrier of information transport.

Magnonics belong to a class of devices based on spin waves
to carry and process information on the nanoscale.

They are the magnetic counterpart of plasmons that are used in nanophotonics
since the wavelength $\lambda$ of magnons is orders of magnitude shorter 
than that of EM waves (photons) of the same frequency. Thus magnonic 
devices represent a serious step toward the fabrication of nanometer-scale 
microwave devices. 

In nanophotonics surface plasmons are used to mediate EM propagation 
in order to beat the "diffraction limit~\cite{Gramotnev}"
requiring the photon wavelength $\lambda < a$ where $a$ is the optical fiber
diameter. If $\lambda$ is on the order of a micron and $a$
on the order of a nanometer, no propagation is possible. The situation
is worse in nano-scale microwave circuits since frequencies are in the GHz
($\lambda$ about 30 cm) and $a$ on the order of a nanometer.

This work is concentrated on electrical means to control
magnonic and spintronic devices exploiting spin-waves and 
spin-currents for signal processing, logic, memories and telecommunication.

Dealing with such magnetic devices implies the requirement of controlling
adequately magnetic fields given the fact that in the microelectronics industry, 
electric voltage (or field) and not electric current control prevails.

Thus, fabrication of practical devices aims at the goal of replacing magnetic field 
control by an electrical one (whenever possible) since
stray magnetic fields might interfere with device operation besides, technically
it is difficult presently to produce large magnetic fields over a nanometer length. 

This work is organised as follows. In section 2 electronic control is
discussed from the general viewpoint of either replacing magnetic control altogether 
or combining electricity and magnetism by either coupling or interconverting 
electrical and magnetic degrees 
of freedom using specific mechanisms with special types of materials such as 
composites, metamaterials or single-phase multiferroics. 
Section 3 is concerned with electronic control of 
magnonic devices whereas section 4 deals with electronic control 
of spintronic devices. Conclusions and outlook are presented in section 5.

\section{Combining electricity and magnetism}

\subsection{Coupling electrical and magnetic degrees of freedom}
Electric and magnetic degrees of freedom when coupled can be described with  
the following free energy expansion~\cite{Eerenstein}:
\bea
- F({\bm E},{\bm H}) = +\frac{1}{2} \epsilon_0 \epsilon_{ij} E_i E_j + \frac{1}{2} \mu_0 \mu_{ij} H_i H_j  \nonumber \\
+ \alpha_{ij} E_i H_j + \frac{1}{2} \beta_{ijk} E_i H_j H_k  +\frac{1}{2} \gamma_{ijk} H_i E_j E_k...
\eea

$\alpha_{ij}$ is first order (in ${\bm E},{\bm H}$ 
components) coupling constant whereas  $\beta_{ijk}$ is first order in ${\bm E}$
second order in ${\bm H}$ whereas $\gamma_{ijk}$ is a first order in ${\bm H}$
second order in ${\bm E}$ components.

$\epsilon_0$ and $\mu_0$ are respectively permittivity and permeability of free space
whereas  $\epsilon_{ij}$ and $\mu_{ij}$ are respectively 
relative permittivity and permeability.

$F$ must satisfy symmetry~\cite{Eerenstein} considerations such as space-inversion symmetry
(satisfied in ferromagnets) and time-inversion symmetry (satisfied in ferroelectrics)
imposing constraints on the expansion coupling constants.
If a multiferroic is both ferroelectric and ferromagnetic, both symmetries are not 
required~\cite{Eerenstein}.
 
Ferroelectric polarization ${\bm P}$ and magnetization ${\bm M}$ responses are 
obtained by differentiating $F$ such as:
\bea
P_i={\left[\frac{\partial F({\bm E},{\bm H})}{\partial E_i}\right]}_{E_i=0}=
\alpha_{ij} H_j +\frac{1}{2} \beta_{ijk} H_j H_k+... \nonumber \\
\mu_0 M_i={\left[\frac{\partial F({\bm E},{\bm H})}{\partial H_i}\right]}_{H_i=0}=
\alpha_{ji} E_j + \frac{1}{2} \gamma_{ijk} E_j E_k+... \nonumber \\
\eea

showing how electric degrees of freedom affect magnetic ones and vice versa.

Recall that electric polarization can also be induced in non-uniformly magnetized materials such that 
${\bm P}~\propto~[({\bm M} \cdot {\bm \nabla}) {\bm M} - {\bm M} ({\bm \nabla} \cdot {\bm M})] $.

Materials containing coupling terms of that nature are magneto-electric (ME) materials
whereas a more general class of materials embodying additional coupling with elastic terms
are called multiferroic which might be composite or single phase.

In composite materials, coupling between magnetic and electrical degrees of 
freedom effect or ME coupling is mediated through elastic interaction such as between 
a magnetostrictive and an electrostrictive (or piezoelectric) substance.

Composite materials made from ferroelectric (or piezoelectric) elements 
containing lower dimensional 
magnetic elements such as thin films or multilayers (laminar coupling), wires (fiber or rod
coupling) and beads (spherical inclusions)~\cite{Bichurin} interacting with 
each other, may be the simplest structures to couple electrical and magnetic 
degrees of freedom.

In addition to ME composites and multiferroic materials, 
dilute magnetic semiconductors (DMS~\cite{Morkoc3}) respond to magnetic fields through 
dispersed magnetic elements that sense the perturbing magnetic field.
Magnetic superlattices (also called metamaterials), 
the magnetic analog of semiconducting superlattices or photonic structures (with 
spatially variable dielectric constant) 
contain spatially modulated magnetization, anisotropy or magnetic phase 
(ferro, antiferro, ferri...) that can 
alter magnon properties such as dispersion relations (gap and group velocity) 
allowing spatially dependent adaptive control. 
 
\subsection{Interconverting electrical and magnetic degrees of freedom}
A magnetic insulator such as YIG (Ferrimagnetic Yttrium Iron Garnet Y$_3$Fe$_5$O$_{12}$~\cite{Morkoc1,Morkoc2}) 
covered with a noble metal such as Pt can interconvert electrical and magnetic degrees of freedom.
A spin current in YIG can be generated (see next section) 
and detected electrically by using spin and charge current interaction~\cite{Kajiwara}. 
The signal can travel over a long distance~\cite{Morkoc1,Morkoc2} in YIG since it 
is an insulator devoid of free charges that can act as scattering sources, moreover it
has a very low intrinsic magnetic damping coefficient (see Table~\ref{tab2}) 
allowing a spin-wave to travel freely without any loss.

Generally, spin currents belong to three types~\cite{Wu}:
\begin{enumerate}
\item SPC (spin-polarized current) made of free (s-type) spin-polarized carriers,
\item SWC (spin-wave current) carried by undulating localized spins (d-type), 
\item STT (spin-transfer torque) current stemming from s-d (double) exchange between
free and localized carriers~\cite{Slonczewski99}.
\end{enumerate}

Current densities ought to be very large (in STT they are about 10$^6$-10$^7$ A/cm$^2$) 
in order to produce magnetic switching, this is why the ultimate 
goal is to rather aim for  a small voltage~\cite{Liu} (electric field control~\cite{Nath})
in order to generate a magnetic field or to produce a magnetic effect since 
it is the conventional control
used in traditional microelectronics and because it is  spatially localized in contrast with
current control that might lead to uncontrolled stray fields.

Magnetic-Electric interconversion between YIG and Pt is based on transfer 
of spin-angular momentum from (localized) magnetization-precession motion (in YIG) 
to (free) conduction-electron spins (in Pt) and spin transfer torque (STT). 
The latter is the reverse process i.e. the transfer of angular momentum from 
conduction-electron spin (free) back to (localized) magnetization. Many 
interconversion aspects are
allowed by the special properties of YIG displayed in Table~\ref{tab2}. 

\vspace{0.5cm}

\begin{center}
\begin{table}[htbp]
\begin{tabular}{l|c}
\hline
Parameter  &               Value       \\        
\hline                                                             
Lattice constant   &         12.376 $\pm$ 0.004 \AA   \\   
Density                                  &      5.17 g/cm$^{3}$     \\         
Band gap                                 &      2.85 eV             \\  
Saturation induction 4$\pi M_S$        &      1750 G              \\  
Cubic anisotropy constant $K_1$          &          -610 J/m$^{3}$   \\  
Anisotropy field $2 K_1/M_S$          &         88 Oe  \\ 
Cubic anisotropy constant $K_2$          &          -26 J/m$^{3}$          \\                       
Curie temperature $T_C$                  &         563 K               \\
Thermal expansion coefficient    &         8.3 x 10$^{-6}$/K  \\                                               
Relative dielectric constant (at 10 GHz)            &       14.7               \\  
Dielectric loss tangent (at 10 GHz)        &       0.0002            \\
FMR linewidth  $\Delta H$ (at 10 GHz) &         0.1 Oe \\    
Intrinsic LLG damping constant  $\alpha $       & 3 x 10$^{-5}$             \\
Land\'e factor    $g$                   &   2.00   \\
\hline
\end{tabular}
\caption{Structural, electric, magnetic, thermal and microwave properties of ferrimagnetic 
Yittrium Iron Garnet~\cite{Morkoc1,Morkoc2} at room temperature in practical units. 
YIG possesses the narrowest FMR (ferromagnetic resonance) 
linewidth of all materials with smallest losses and almost zero LLG (Landau-Lifshitz-Gilbert) damping.}
\label{tab2}
\end{table}
\end{center}

Progress in electronic control of magnetic devices summarized in Table~\ref{tab3}
shows that many magnetic properties can be tightly controlled with an electric
field leading to control of spin-waves and spin-currents as described below.

\begin{widetext}

\begin{table}[h!]
\begin{tabular}
{|p{162pt}|p{162pt}|p{162pt}|}
\hline
Electrically controlled quantity & 
Comments& 
Reference \\
\hline
Exchange interaction &
Proposal for electric field  modification of local exchange between neighbouring magnets & 
Gorelik \etal~\cite{Gorelik} (2003).\\
\hline
Nature of magnetic phase & 
Ferromagnetic ordering in hexagonal HoMnO$_3$ is reversibly controlled by an electric field & 
Lottermoser \etal~\cite{Lottermoser} (2004). \\
\hline
Coercivity &
Electric control of coercivity in CoFeB/MgO/CoFeB magnetic tunnel junction & 
Wang \etal~\cite{Wang} (2005).\\
\hline
Exchange bias &
Electric-field control of exchange bias in multiferroic epitaxial heterostructures &
Laukhin \etal~\cite{Laukhin} (2006). \\
\hline
Sensors, transducers and microwave devices &
ME control in composites made with magnetostrictive and piezoelectric elements &
Nan \etal~\cite{Bichurin} (2008). \\
\hline
Magnetic domain wall motion and magnetization direction &
Writing/Erasure of ferromagnetic domains and electric control of domain wall motion &
Lahtinen \etal~\cite{Lahtinen} (2012). \\ 
\hline
Anisotropy &
Electrically induced large magnetization reversal in multiferroic 
Ba$_{0.5}$Sr$_{1.5}$Zn$_{2}$(Fe$_{0.92}$Al$_{0.08}$)$_{12}$O$_{22}$ &          
Chai \etal~\cite{Chai} (2014).\\
\hline
\end{tabular}
\caption{Selected progress milestones in electronic control of magnetic devices.}
\label{tab3}
\end{table}

\end{widetext}

\section{Electronic control of spin-waves in thin film devices}

Spin waves can be  generally divided into three categories~\cite{Lenk}: \\

a- Magnetostatic spin waves (MSW) originating from long-range dipolar interactions 
between elements whose typical size is the micron. Damon-Eshbach~\cite{Damon}
MSW modes are transverse (the wavevector ${\bm k}$ is perpendicular to 
local magnetization ${\bm M}$). \\
Magnetostatic Surface Spin-Waves (MSSW), the magnetic counterpart of Surface Acoustic 
Waves (SAW) can also be excited in thin films or stripes made of YIG and 
their energy is on the order of a few GHz~\cite{Lenk}. \\
b- Exchange spin-waves (ESW) originating from short-range Heisenberg exchange interactions 
between elements whose typical size is the nanometer. Their energy is
on the order of a tenth of a GHz (or a $\mu$eV) reaching the THz~\cite{Lenk} 
in antiferromagnets.\\
c- Dipolar-Exchange Spin Waves (DESW) in the case of devices of mixed length 
type~\cite{Lenk}, such as Magnetic Quantum Dots (MQD) laid out periodically as planar arrays, 
with a typical in-plane length (MQD diameter) on the order of a micron and a 
perpendicular-to-plane length (MQD height) on the order of a nanometer.

In the latter case, spin-waves with both types of contributions occur as
predicted for the first time by Herring-Kittel~\cite{Herring} for infinite 
media and by Clogston \etal~\cite{Clogston} for finite media such as 
ellipsoids.

\begin{widetext}

\begin{table}[h!]
\begin{tabular}
{|p{162pt}|p{162pt}|p{162pt}|}
\hline
Electrical control type & 
Comments& 
Reference \\
\hline
Spin-Wave logic gates &
Current-controlled Mach-Zender type interferometer based on MSSW propagation in YIG thin films &
Kostylev \etal~\cite{Kostylev} (2005). \\
\hline
Spin-Wave phase and wavelength &
Current control of phase and wavelength of Damon-Eshbach MSW propagating in  
Permalloy (Ni$_{81}$Fe$_{19}$) ribbons deposited over Cu stripes &
Demidov \etal~\cite{Demidov} (2009). \\
\hline
Spin-Wave frequency control &
Electric-field tuning of ESW frequency at room temperature using multiferroic BiFeO$_3$ &
Rovillain \etal~\cite{Rovillain} (2010). \\
\hline
Amplification of spin-waves &
Electric-field amplification of ESW in YIG/Pt bilayers using Interfacial Spin Scattering &
Wang \etal~\cite{Wang2} (2011). \\
\hline
Generation of spin-waves &
Electric-field-induced ESW generation using multiferroic ME cells &
Cherepov \etal~\cite{Nath} (2014). \\
\hline
\end{tabular}
\caption{Selected progress milestones in electronic control of magnonic devices with several types
of spin-waves.}
\label{tab4}
\end{table}

\end{widetext}

\subsection{Spin-wave generation energetics}
Heisenberg model for localized spins in a ferromagnet gives an
interaction energy for a pair of neighbouring spins ${\bm S_i},{\bm S_{i+1}}$ 
as $-2 J_{ex} {\bm S_i} \cdot {\bm S_{i+1}}$ where $J_{ex}$ is the exchange integral.
The Curie temperature is obtained from 
$k_B T_c=\frac{2}{3} J_{ex} z S(S+1)$ where $z$ is the coordination number
and $S$ the spin value.
$J_{ex}$ leads also to spin-wave energy dispersion: $\hbar \omega_k=4 J_{ex} S (1-\cos k a)$ 
where $k$ is the wave vector.
Considering a 1D array with lattice parameter $a$, reversing a single spin 
costs the spin-flip energy  $4 J_{ex} S^2$.  \\
For a set of $n$ spins, the total 
energy is $4n J_{ex} S^2$ meaning that propagating a spin-flip across a 
distance $na$ requires such energy. 
By comparison, spin-wave propagation can be performed with a wavelength 
chosen to match the same distance $\lambda=na$. The corresponding
energy can be evaluated for the wavevector $k=2\pi/na$ substituted in $4 J_{ex} S (1-\cos k a)$.
This gives the energy $2 J_{ex} S k^2 a^2$ equal to  $8 J_{ex} S \pi^2/n^2 $  in the limit $ka=2\pi/n << 1$. \\
As an application, we consider a Ni ribbon as part of a Ni/Permalloy (Ni$_{81}$Fe$_{19}$)
bilayer considered as a spin-wave bus (see next subsection) to propagate spin-waves. 
Despite the fact, Ni is an itinerant magnet where the Heisenberg picture does 
not strictly apply, we infer that $J_{ex}=1.45 \times 10^{-21}$ Joules or 9 meV 
(in comparison, Stoner exchange integral is 1.01 eV~\cite{Coey}) from the above Curie 
temperature formula with $S=\frac{1}{2}$, $T_c=629$K and coordination number 
$z=12$ (from Ni FCC structure). \\
As an example, the spin-flip propagation energy for 1000 spins is $1.45 \times 10^{-21}$ 
Joules whereas the corresponding spin-wave energy (with Ni lattice parameter $a$=3.52 \AA) 
is $5.71 \times 10^{-26}$ Joules which is about five orders of magnitude smaller. \\
The spin-wave to spin-flip energy propagation ratio is independent of the value
of $J_{ex}$ and strongly decreases with the number $n$
of spins as $\frac{2 \pi^2}{S n^3}$ suggesting that spin-wave propagation is definitely
lower in terms of energy cost.  

\subsection{Voltage-induced spin wave generation with hybrid cells}
Single-phase multiferroics have in general small ME coupling. 
Instead of using large consumption devices such as inductive antennas or 
STT currents in order to generate spin-waves,  
Cherepov \etal~\cite{Nath} used hybrid (multiferroic ME) cells 
consisting of a magnetostrictive Ni layer and a piezoelectric 
substrate PMN-PT i.e. {[Pb(Mg$_{1/3}$Nb$_{2/3}$)O$_{3}$]}$_{(1-x)}$~-~{[PbTiO$_3$]}$_x$. \\
PMN-PT (lead magnesium niobate-lead titanate) is a ferroelectric 
relaxor~\cite{Liu} with a large relative dielectric constant 
$\epsilon_r$. For $0< x < 0.35$, 
the electromechanical coupling and piezoelectric
coefficients of PMN-PT are very large making it a sensitive material for ME control.
  
Applying an AC voltage to (PMN-PT) induces an alternating strain
in the piezoelectric material. The strain transmitted to the 
magnetostrictive Ni layer produces local anisotropy variation 
resulting in easy axis reorientation that pulls on the magnetization.
Magnetization oscillations propagate in the form of spin waves
in a Ni/Permalloy bilayer lithographically 
shaped in the form of a stripe that is called a spin-wave bus. 
While the Ni layer provides the desired magnetostriction,
NiFe being a soft magnetic material is favorable to spin-wave propagation because of
its low LLG damping constant $\alpha$. 

In order to describe spin-wave generation electronically, we
need a process that entails tilting a single spin belonging
to an ordered spin array. This entails application of a localized 
magnetic induction field $B_{ex}$ for a short time. 

The value of $B_{ex}$ required to tilt a spin in the ferromagnetic 
stripe should be on the order of $B_{ex}=B_{dem} \tan\theta $ 
with $\theta$ the tilt angle and $B_{dem}$  the demagnetization field  
orthogonal to $B_{ex}$. This originates from the fact the internal field sensed by 
any spin $M$ is the sum of the external field $B_{ex}$ and the demagnetization 
field $B_{dem}$ with $B_{dem}=-4\pi N M$. $N$ is the demagnetization coefficient and $M$ 
the local magnetization.\\
If we approximate the ferromagnetic stripe by a saturated thin film, the 
demagnetization coefficient $N=1$ and $M=M_s$. Thus we infer that the required 
excitation field is $B_{ex}=4 \pi M_s \tan\theta$.

This field can be generated by an antenna~\cite{Nath} delivering a current pulse 
or an ME capacitive cell providing voltage pulse 
whose duration is determined by the operating frequency $f_{LO}$ of the device local 
oscillator.

At a distance $r$, Amp\`ere law provides the expression $H_{ex}=\frac{I}{2\pi r}$ relating
the required current intensity $I$ to the required magnetic field $H_{ex}=B_{ex}/\mu_0$. 
Thus the required energy is $\frac{R I^2}{f_{LO}}$ where $R$ is the resistance of the antenna traversed by $I$. \\  

As an example, we take $\theta=1$\deg, $M_s$=485 emu/cm$^3$ 
(Ni saturation magnetization) and $f_{LO}$=5 GHz. 
At a nominal distance $r$ about 10 times the minimal feature (14 nm), 
we get the required current $I$=7.48 mA.

Taking antenna typical dimensions such as 
length, height and width $\ell,a, b$ all around 10 times the minimal feature value
and a standard metallic resistivity of $\rho$=1 $\mu\Omega$.cm, we get
antenna resistance $R=\frac{\rho \ell}{ab} \approx 0.07 \Omega$ and  
dissipation energy $\frac{R I^2}{f_{LO}} \approx 8 \times 10^{-16}$ Joule. \\

Moving on to the ME capacitive cell made with a material 
whose ME coefficient $\beta= \frac{\delta H}{\delta E}$ is not small, 
we may substitute efficiently an electric field to a magnetic 
induction field via $E_{ex}=\frac{B_{ex}}{\mu_0 \beta}$. \\
Thus the voltage required for tilting a spin is $V_{ex}=E_{ex} \ell$ where $\ell$
is the inter-plate distance of a cell whose capacitance $C=\epsilon_r \epsilon_0 ab/\ell$
with energy $\frac{1}{2} C V^2$. \\

A material such as Fe$_3$O$_4$/PMN-PT has an ME coefficient $\beta=67$ Oe.cm/kV 
(see Liu \etal~\cite{Liu}) and a relative dielectric constant around several 1000. 

Taking all dimensions such as length, height and width 
$\ell,a,b$ about 10 times the minimal feature value, we get a capacitance
$C=\frac{\epsilon_r \epsilon_0 ab}{\ell}$=1.23 $\times 10^{-15}$ F, a
voltage of 22.2 mV and a tilting energy of $\frac{1}{2} C V^2$=3 $\times 10^{-15}$ Joule
which is three orders of magnitude below the antenna case.

The capacitive over resistive (antenna) energy ratio  
$\eta= \frac{1}{2} C V^2/\frac{R I^2}{f_{LO}}$ can
be expressed in a scaling form as 
$\eta=\frac{\epsilon_0 \epsilon_r \ell^2 f_{LO}}{8 \pi^2 \rho \beta^2}$
where $\ell$ replaces all lengths in the device implying that as 
minimal feature continues to decrease $\eta \sim \ell^2$ will decrease. \\
Moreover $\eta \sim \frac{\epsilon_r}{\beta^2}$ implying
that we need multiferroic materials with a smaller $\epsilon_r$ and a 
larger ME coefficient $\beta$
in order to keep $\eta$ decreasing, in contrast to materials such as 
PMN-PT with both ($\epsilon_r$,$\beta$) large.

Once the spin is tilted, its time variation follows space-dependent LLG equation:

\be
 \frac{\partial {\bm M({\bm r},t)}}{\partial t} =  -\gamma {\bm M({\bm r},t)}\times {\bm H}
-\frac {\alpha \gamma}{|{\bm M}|} {\bm M({\bm r},t)} \times [{\bm M({\bm r},t)} \times {\bm H}]
\ee

where $\gamma $ is the gyromagnetic ratio, ${\bm M}$ the magnetization vector, 
${\bm H}$ the effective field obtained from total energy and $\alpha$ 
the intrinsic damping parameter (see Table~\ref{tab2}). LLG equation describes a propagating 
Bloch equation for a moment precessing around magnetic field  ${\bm H}$ direction 
and damped by the $\alpha$ term that forces the moment to precess closer to the magnetic field direction 
thus reducing the initial tilt angle $\theta$. Choosing materials such as Ni, Ni/NiFe
bilayers as well as YIG that possess small $\alpha$, allows long-distance signal propagation 
since $\theta$ diminishes very weakly in the course of time.

\section{Electronic control of spintronic devices}

\subsection{Issues in electric-field controlled spintronic devices}

Substituting magnetic field control by an electrical one is possible in semiconductors 
via spin-orbit interaction since it allows generation and manipulation of carrier spins
by an electric field~\cite{Samarth}. 

Spin-orbit interaction has also been shown to allow electric
control of spin-waves in single-crystal YIG waveguides which
paves the way to develop electrically tunable magnonic 
devices~\cite{Vignale,Flatte}.

Another alternative is a tunable spin current that allows to generate a magnetic field or 
produce a magnetic effect (such as reversal or alteration of magnetization)  
since a spin current targets interconversion between charge and spin degrees 
of freedom~\cite{Wu,Dietl}.

In analogy with ordinary electronics, spintronics is based on several operations
such as spin injection, filtering, accumulation, detection and pumping~\cite{Samarth}.

The notion of spin coherence underlies all these operations, that is preservation
of a given spin state over long distances despite the presence of impurities,
dislocations, noise, stray magnetic fields, Earth magnetic field etc...

Spin injection might be done optically in complete analogy with 
Haynes-Shockley experiment. It may also be done with carbon nanotubes 
since they do not alter the spin state over 
large distances (see Dietl \etal~\cite{Dietl}).

Spin filtering is essentially the separation of spin-polarised carriers which is
required for avoiding spin-flips that alter carrier spin states akin to geminate
recombination between photo-generated electron-hole pairs.
Thin ferromagnetic layers as in spin valves (see Dietl \etal~\cite{Dietl}) and
chiral materials such as monolayers of double-stranded DNA molecules can be 
used~\cite{Gohler}.

Spin accumulation is concerned with increase of concentration of spin polarized 
carriers without destroying their coherence or inducing spin-flips among them
whereas spin detection relates to non-destructive determination of spin value.

Spin pumping~\cite{Tserkovnyak} occurs in Ferromagnetic-Normal bilayers, 
when the precessing magnetization in the Ferromagnet (F) injects a spin-current 
into a normal metal (N) through the F-N interface (as in the YIG/Pt case described
previously).

\begin{figure}[htbp]
  \begin{center}
    \begin{tabular}{c}
      \resizebox{50mm}{!}{\includegraphics{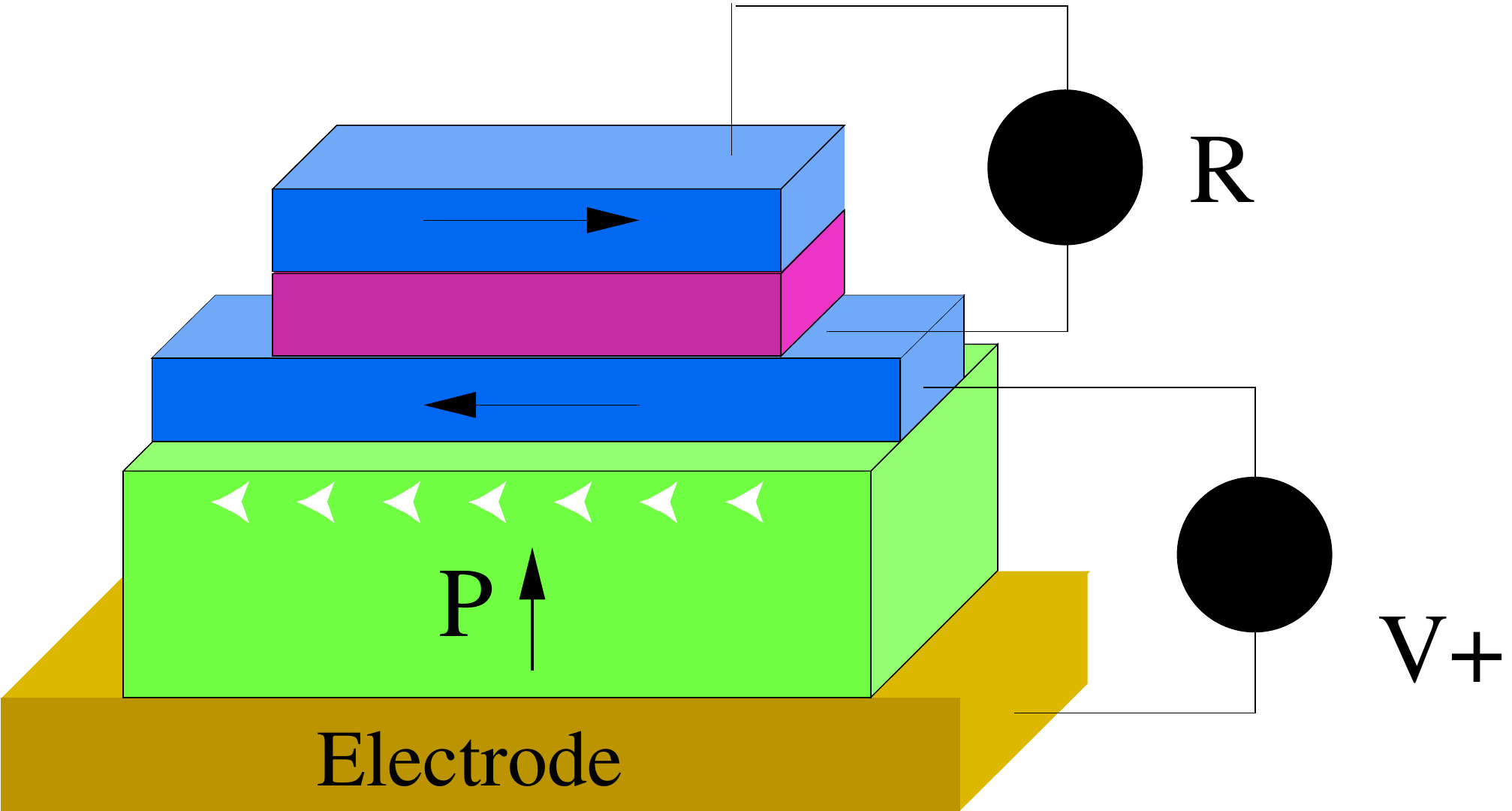}} \\
      \resizebox{50mm}{!}{\includegraphics{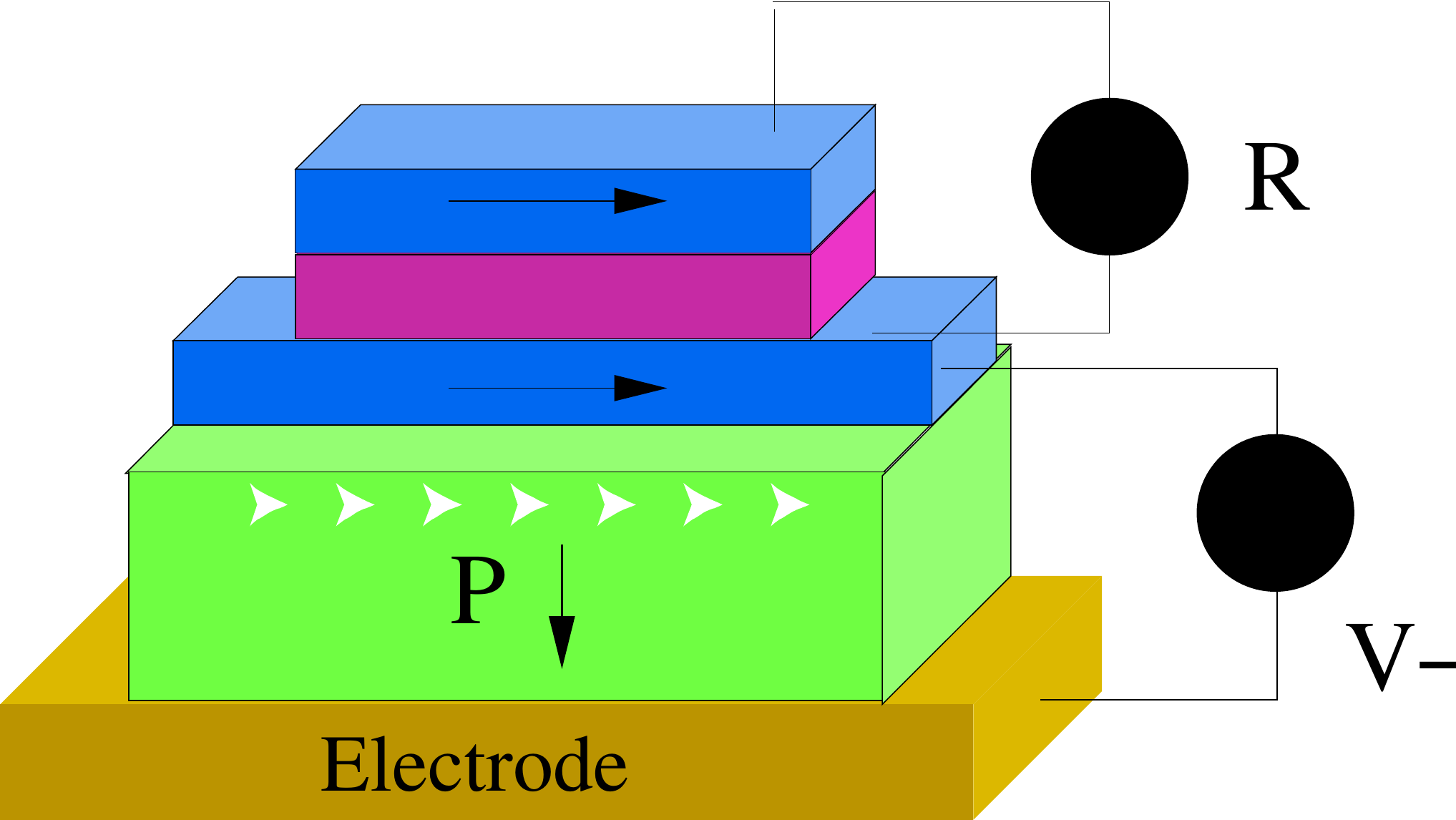}} \\ 
    \end{tabular}
    \caption{(Color on-line) MeRAM cell showing the electric field control and the 
different materials composing the memory cell. A spin-valve made from two blue layers (ferromagnets) 
separated by a metallic spacer (in magenta) has its resistance {\bf R} monitored.
The upper blue layer made from a hard magnetic material is free and the lower 
blue layer is made from a soft magnetic material in close contact with  
a (green layer) material made from a ferroelectric antiferromagnet, a single-phase multiferroic. 
Top figure shows positive voltage {\bf V} applied to electrode (gold yellow) with blue layer
magnetizations (black arrows) antiparallel resulting in a high resistance {\bf R}. 
Bottom figure shows negative voltage {\bf V} applied while blue layers have parallel 
magnetizations yielding a very low resistance {\bf R}. 
Magnetization (black arrow) in each lower blue layer follows green material 
magnetization represented by small white arrows.
In a multiferroic, polarization {\bf P} and magnetization are coupled~\cite{Zhao}. 
Thus when  {\bf P} is up ({\bf V} positive),
magnetization (represented by small white arrows)  is oriented to the left and when {\bf P} is 
down ({\bf V} negative), magnetization is oriented to the right. 
Adapted from Bibes \etal~\cite{Bibes}}
    \label{cell}
  \end{center}
\end{figure}

Many of the spin manipulation processes described above intervene in the progress milestones
for electric-field control of spintronic devices displayed in Table~\ref{tab5}.

\begin{widetext}

\begin{table}[htbp]
\begin{tabular}
{|p{162pt}|p{162pt}|p{162pt}|}
\hline
Electrical control type & 
Comments& 
Reference \\
\hline
Driving of microwave oscillation & 
GHz oscillator excited by STT current& 
Kiselev \etal~\cite{Kiselev} (2003). \\
\hline
Magnetic Random-Access Memories (MRAM) & 
Electric control of CoFeB/MgO/CoFeB tunnel magnetoresistance switching & 
Wang \etal~\cite{Wang} (2005). \\
\hline
Spin transport in nanotubes &
Electric-field control of magnetoresistance in carbon nanotubes connected by ferromagnetic leads &
Sahoo \etal~\cite{Sahoo} (2005). \\
\hline
Spin transport in Silicon &
Ballistic spin-dependent hot-electron filtering through ferromagnetic thin films &
Appelbaum \etal~\cite{Appelbaum} (2007). \\
\hline
Memory Read-Write operations & 
Information processing in MRAM cell made with a DMS: (Ga,Mn)As & 
Mark \etal~\cite{Mark} (2011). \\
\hline
Spintronic logic gates & 
Voltage-controlled spin selection and tuning in graphene nanoribbons for logic &
Zhang~\cite{Zhang} (2014). \\
\hline
\end{tabular}
\caption{Selected progress milestones in electronic control of spintronic devices}
\label{tab5}
\end{table}

\end{widetext}

\subsection{Electric-field controlled magneto-electric RAM devices}
Magnetic Random-Access Memories (MRAM) abolish the distinction between volatile 
storage (used during processing) and permanent massive storage.
A RAM is used by a CPU or a DSP (Digital Signal Processor) during processing,
for loading an operating system (OS) or enabling application programs (AP). 
The implication that the OS and the AP are permanently loaded in MRAM brings
a paradigm shift in computing that has far reaching consequences in terms of 
computing speed and efficiency. The initial attempts to design 
MRAM cells were based on domain wall motion control with a strong electric 
current (racetrack-type memories~\cite{Dietl}). \\
MeRAM memory is a new type of voltage controlled RAM based on ME multiferroic interacting
with a ferromagnet. The multiferroic is a ferroelectric antiferromagnet
whose electric polarization induces an internal magnetization (see fig.~\ref{cell})
that controls the neighbouring ferromagnet magnetization at the interface. 
An example ferroelectric antiferromagnet is BiFeO$_3$ (BFO) that displays both 
ferroelectricity and antiferromagnetic order at room temperature~\cite{Talbayev}. \\  

BFO has a rhombohedral perovskite crystallographic structure.
It is a high-temperature ferroelectric (with Curie temperature $T_c \approx$ 1100 K) 
possessing a large ferroelectric dipole moment $\approx$ 100 $\mu$C/cm$^2$. 
At room temperature, bulk crystalline BFO is antiferromagnetic~\cite{Talbayev}
with N\'eel temperature T$_N$ of 640 K. 
Voltage control of BFO magnetic state has been shown both in bulk and in 
thin film case making it an excellent candidate for ME applications.

The ferroelectric polarization and the antiferromagnetic vector in BFO 
are coupled~\cite{Zhao} in a way such that by reversing the polarization, 
the antiferromagnetic spins rotate. Nevertheless the antiferromagnetic structure of BFO
is complicated~\cite{Talbayev} making the work of Chu \etal~\cite{Chu} more appealing. 

The latter~\cite{Chu} have shown that micrometre-size ferromagnetic CoFe dots deposited 
on a BFO film are consistently coupled, in a reversible manner, with the 
BFO antiferromagnetic spins. 
This implies that when an in-plane electric field is applied, 
CoFe dot magnetization 
rotates by 90\deg and when voltage polarity is reversed
the original CoFe magnetic state is retrieved.

\section{Conclusions and Outlook}
Electric control of magnonic and spintronic devices is steadily progressing~\cite{Liu}, 
offering lower energy consumption devices and paving the way to potentially 
solve the nagging interconnection delay problem. \\

Simultaneously, femtosecond optical control is also progressing in antiferromagnets 
that represent the largest class of spin ordered materials in Nature with
spin-wave excitations occurring typically at frequencies as high as a THz.
Eventually, this will lead to extremely fast control~\cite{Kampfrath} of magnetic 
devices to reach the THz regime unlocking the frequency stalling problem 
around a few GHz in present CMOS devices. 

Kampfrath \etal~\cite{Kampfrath} used optical femtosecond
pulses to control spin-waves in antiferromagnetic NiO and more recently
Shuvaev \etal~\cite{Shuvaev} demonstrated electrical control of a 
dynamic ME effect in DyMnO$_3$, a single-phase multiferroic material. 
ME coupling with spin-waves leads to rotation of polarization plane 
of light propagating across a sample and a static voltage allows to control 
light amplitude and polarization plane rotation.  

The THz barrier is the last one to overcome among four making
the pillars of Digital Technology that is continuously thriving toward 
THz operation speed, Teraflop processing, Terabyte storage and finally 
Terabit/sec communication.

\end{document}